# Enhanced superconductivity accompanying a Lifshitz transition in electron-doped FeSe monolayer

X. Shi[1], Z.-Q. Han[1,2], X.-L. Peng[1], P. Richard[1,3], T. Qian[1,3], X.-X. Wu[1], M.-W. Qiu[1], S.C. Wang[2], J.P. Hu[1,3], Y.-J. Sun[1*] & H. Ding[1,3*]

[1]Beijing National Laboratory for Condensed Matter Physics, and Institute of Physics, Chinese Academy of Sciences, Beijing 100190, China

[2]Department of Physics, Renmin University, Beijing, 100872, China

[3]Collaborative Innovation Center of Quantum Matter, Beijing 100190, China

**The origin of enhanced superconductivity over 50 K [1, 2, 3, 4] in the recently discovered FeSe monolayer films grown on SrTiO₃ (STO), as compared to 8 K in bulk FeSe [5], is intensely debated. As with the ferrochalcogenides $A_x Fe_{2-y} Se_2$ [6, 7] and potassium doped FeSe [8, 9], which also have a relatively high superconducting critical temperature ($T_c$), the Fermi surface (FS) of the FeSe/STO monolayer films is free of hole-like FS, suggesting that a Lifshitz transition by which these hole FSs vanish may help increasing $T_c$. However, the fundamental reasons explaining this increase of $T_c$ remain unclear. Here we report a 15 K jump of $T_c$ accompanying a second Lifshitz transition, characterized by the emergence of an electron pocket at the Brillouin zone (BZ) centre, that is triggered by high electron doping following *in-situ* deposition of potassium on FeSe/STO monolayer films. Our results suggest that the pairing interactions are orbital-dependent with the $d_{xy}$ orbital playing a determining role in generating enhanced superconductivity in FeSe.**

Until now, the highest $T_c$ among all iron-based superconductors is achieved in FeSe monolayer films [3, 4, 10, 11]. The exact mechanism of this superconductivity enhancement in these systems, as well as in other FeSe-based materials such as $A_x Fe_{2-y} Se_2$ [6] and (Li,Fe)OHFeSe [12], has become a central focus in iron-based superconductivity. FeSe-based materials with relatively high $T_c$'s share one common key point in their FS topology: the absence of hole pockets at the BZ centre [7, 13, 14]. The importance of this FS topology to superconductivity has been further supported by doping electron carriers on the surface of thick FeSe films or crystals using potassium deposition [8, 9], or in their bulk using liquid-gating technique [15]. In this context, it is natural to ask up to what level the monolayer FeSe/STO can be electron-doped and how superconductivity is linked to the fermiology at high electron doping.

In this paper we perform high-resolution angle-resolved photoemission spectroscopy (ARPES) measurements on electron-doped FeSe/STO monolayers. Starting from a well-annealed sample which originally possesses a relatively high electron concentration, as confirmed by large electron FS pockets (see Fig. 1a), we deposit K atoms *in-situ* onto the surface and achieve a higher doping level.

As determined in previous ARPES studies [2, 3] and in our current experiment, the FS topology of FeSe/STO(001) monolayer films shown in Fig. 1a consists of nearly doubly-degenerate electron-like pockets centred at the M point, in contrast to FeSe bulk crystals [16] and most of the ferropnictide superconductors [17]. We then deposit potassium (K) onto the surface of the film and check the evolution of the FS. Figure 1b shows the FS map after evaporating a small dose of K. The area of electron pocket at M increases from ~ 8.2% of the BZ in the pristine sample (Figure 1a) to ~ 10.4%, indicating that K atoms introduce extra electron carriers into the system. However, further deposition of a similar dose of K does not introduce as many electrons as the first time, and the electron carrier concentration of the system tends to slowly saturate (~ 10.7% of the BZ after the third round of K deposition).

Surprisingly, the FSs obtained after more than one round of deposition exhibit strong intensity at Γ. This is clearly confirmed by the momentum distribution curves (MDCs) shown in Fig. 1e. Such a change suggests that the system evolves towards a Lifshitz transition, possibly caused by a chemical potential shift. Figure 1f plots the FS evolution of the monolayer FeSe/STO upon K coating, which is more complicated than for the reported results on FeSe thick film [8, 9].

To understand where the intensity at Γ originates from, we investigate carefully the low-energy electronic structure in detail. We show in Figs. 2a,b the band structure near Γ and M, respectively, along the cuts indicated in the inset of Fig. 1f. The hole-like bands around Γ and the electron-like bands around M shift toward high binding energy, which is consistent with the expected electron doping by K atoms. We note that a simple rigid chemical potential shift cannot describe the band structure evolution, like in the case of K doped FeSe thick films [9]. Figure 2c shows the spectra recorded at 70 K after division by the FD function convoluted by the resolution function. While no band is observed in the pristine sample in the measured range above $E_F$ at Γ, an additional electron-like band possibly crossing $E_F$ appears for the $x =$ 0.212 and 0.214 samples on the unoccupied side of the spectrum (See also the EDCs in Fig. S1b). Scanning tunnelling microscopy (STM) measurements reveal that the bottom of this band locates at 75 meV above $E_F$ in the pristine monolayer [18]. By comparing results using the He Iα (21.2 eV) and He II (40.8 eV) lines of a He discharge lamp (Fig. 2f), and taking advantage of the opposite behaviour of the photoemission cross section of Fe $3d$ and Se $4p$ in this energy range (Fig. 2f) [19], we conclude that this electron band has a dominant Se $4p$ orbital character. Band calculations [18, 20] and previous ARPES studies on similar materials [21] demonstrate that the Se $4p_z$ orbital is hybridized with the Fe $3d_{xy}$ orbital at Γ. The position of this band is quite sensitive to the Se height (on the very top of the film) [18, 20] and this might be responsible for the relatively large energy shift upon K deposition on the surface.

We then check the superconductivity of the samples. Following a standard procedure, we show in Fig. 3a the temperature dependence of the symmetrized EDCs at $k_F$ near the M point for the pristine FeSe/STO monolayer. In agreement with previous ARPES results [2, 3, 22, 23], the FS is clearly gapped at low temperatures. We fit the experimental data with a phenomenological model for the superconducting gap [24], and display the extracted results in Fig. 3d. The gap size is about 10 meV and closes at around 55 K, which is comparable to reported values [2, 3, 22, 23]. Similarly, we show symmetrized EDCs in Figs. 3b-c for electron doping levels $x$ = 0.208 and 0.212, respectively, after K deposition. The corresponding fitting results are displayed in Figs. 3e-f. For $x$ = 0.208, the gap size and the $T_c$ does not change much. Interestingly, the gap size jumps to ∼ 15 meV and the closing temperature increases to ∼ 70 K after further K doping to $x$ = 0.212. We have checked that these values are almost unchanged with further doping.

The gap that we observe is symmetric in all the low temperature spectra (see Figs. S2 of the supplementary material), which is a characteristic feature of the superconducting gap, in contrast to the high temperature ones. Moreover, the symmetrized EDCs for $x$ = 0.212 (Fig. 3c) show the same spectral weight transfer or filling behaviour as with the pristine monolayer FeSe/STO. We conclude that the system evolves into an enhanced superconducting state upon K coating, with the "transition" point at $x$ = 0.212 corresponds exactly to the appearance of the pronounced intensity at Γ in the FS.

We now further investigate the enhanced superconductivity of the $x$ = 0.212 sample in the momentum space. Figure 3i displays a series of symmetrized EDCs at various $k_F$ points, as indicated in Fig. 3g. The fitting results plotted in a polar representation in Fig. 3j show that the superconducting gap around the M point is isotropic within our experimental uncertainties. Since new electronic states appear at Γ near $E_F$, we checked the temperature dependence of the EDCs at this point. The EDCs were divided by the FD function and displayed in Fig. 3h. There is also a gap feature here with a similar size of 15 meV as that around M.

We summarize our results in Fig. 4. The data of bulk FeSe, including K-doped thick FeSe films [9] and liquid-gated FeSe thin flakes [15], are also plotted for comparison. We note that the electron concentration of FeSe under gating was set based on the gate voltage and the FS evolution of bulk FeSe [9, 15]. The FeSe system undergoes two Lifshitz transitions upon electron doping and the three typical FS topologies are sketched in the Figure. Superconductivity is suddenly enhanced at each transition. Based on our knowledge of the first Lifshitz transition, during which the FS pockets around $\Gamma$ vanish, one may expect a suppression of $T_c$ once a FS pocket appears again at the BZ centre. However, our results revealed the precise opposite behaviour. Although we cannot totally rule out a positive influence of an increase of density-of-states at $E_F$ due to the additional electron pocket, we notice that the $2\Delta/k_B T_c$ ratio varies from $\sim 4.5$ to $\sim 5.1$, in contrast to the expectation for an increase of density-of-states in the BCS framework, for which this ratio should remain constant.

Our observations raise the possible importance of orbital-dependent interactions. Indeed, the $d_{xz}/d_{yz}$ character of the orbitals sinking below $E_F$ across the first Lifshitz transition (accompanying a jump of $T_c$ [9, 15]) is different from the $p_z/d_{xy}$ orbital character emerging at $\Gamma$ across the second Lifshitz transition. Interestingly, in contrast to ARPES measurements on ferropnictide superconductors [27], the outer electron FS pocket at the M point in FeSe/STO, attributed to the $d_{xy}$ orbital, has a larger gap than that of the inner FS pocket according to a recent ARPES study [28]. Such a strong orbital dependence of superconducting pairing interactions has not been obtained in previous theoretical treatments.

There is a possible phenomenological explanation to our observation. In the bulk FeSe, it is known that the presence of the $d_{xz}/d_{yz}$ hole FS pockets strongly favours a nematic electronic state [16, 29, 30]. Thus, their absence can suppress the nematic order to enhance superconductivity, which explains the enhancement at the first Lifshitz transition. In the extended $s$-wave or $s\pm$ pairing scenario, the emergence of the $d_{xy}/p_z$ electron pocket at the second Lifshitz transition can strengthen the $d_{xy}$ intra-orbital pairing, which is consistent with the observation that the gap enhancement is on the pockets attributed to the $d_{xy}$ orbitals. Our results call for a microscopic model involving orbital dependence to explain superconductivity and its enhancement in FeSe/STO. Orbital-dependent AFM interactions must be required in order to understand our results, even qualitatively.

**Methods**

Monolayer films of FeSe were grown on 0.05wt% Nb-doped $SrTiO_3$ substrates after degassing for 2 hours at 600 ℃ and then annealing for 12min at 925 ℃. The substrates were kept at 300 ℃ during the film growth. Fe (99.98%) and Se (99.999%) were co-evaporated from Knudsen cells with a flux ratio of 1:10 (which were measured by a quart crystal balance) and the growth rate of 0.31 UC/min. The growth process was monitored using Refection high-energy electron diffraction (RHEED). After growth, the FeSe monolayer films were annealed at 350 ℃ for 20h (see RHEED image in Fig. S4), and subsequently transferred *in situ* into the ARPES chamber. ARPES measurements were performed at the Institute of Physics, Chinese Academy of Sciences, using a R4000 analyser and a helium discharge lamp, under ultrahigh vacuum better than $3 \times 10^{-11}$ torr. The energy resolution was set to ~ 5 meV for gap measurements and ~ 10 meV for the band structure and FS mapping, while the angular resolution was set to 0.2°. Deposition of the potassium atoms was carried out in the ARPES preparation chamber using a commercial SAES alkali dispenser, during which the samples were kept at low temperature (20 - 30 K).

**References**

1. Wang, Q.-Y. *et al.* Interface-Induced High-Temperature Superconductivity in Single Unit-Cell FeSe Films on $SrTiO_3$. *Chin. Phys. Lett.* **29**, 037402 (2012).

2. Liu, D. F. *et al.* Electronic origin of high-temperature superconduc-tivity in single-layer FeSe superconductor. *Nature Commun.* **3,** 931 (2012).

3. Tan, S. Y. *et al.* Interface-induced superconductivity and strain-dependent spin density waves in $FeSe/SrTiO_3$ thin films. *Nature Mater.* **12,** 634–640 (2013).

4. Zhang, Z. C. *et al.* Onset of the Meissner effect at 65 K in FeSe thin film grown on Nb-doped $SrTiO_3$ substrate . *Sci. Bull.* **60**(14), 1301–1304 (2015).

5. Hsu, F. C. *et al.* Superconductivity in the PbO-type structure FeSe. *Proc. Natl Acad. Sci. USA* **105,** 14262–14264 (2008).

6. Guo, J. G. *et al.* Superconductivity in the iron selenide $K_xFe_2Se_2$ ($0 \leq x \leq 1.0$). *Phys. Rev. B* **82**, 180520(R) (2010).

7. Qian, T. *et al.* Absence of a Holelike Fermi Surface for the Iron-Based $K_{0.8}Fe_{1.7}Se_2$ Superconductor Revealed by Angle-Resolved Photoemission Spectroscopy. *Phys. Rev. Lett.* **106,** 187001 (2011).

8. Miyata, Y. *et al.* High-temperature superconductivity in potassium-coated multilayer FeSe thin films. *Nature Mater.* **14**, 775–779 (2015).

9. Wen, C. H. P. *et al.* Anomalous correlation effects and unique phase diagram of electron-doped FeSe revealed by photoemission spectroscopy. *Nat. Commun.* **7,** 10840 (2016).

10. He, S. L. *et al.* Phase diagram and electronic indication of high-temperature superconductivity at 65 K in single-layer FeSe films. *Nature Mater.* **12**, 605–610 (2013).

11. Ge, J. F. *et al.* Superconductivity above 100 K in single-layer FeSe films on doped $SrTiO_3$. *Nature Mater.* **14**, 285–289 (2014).

12. Liu, X. F. *et al.* Coexistence of superconductivity and antiferromagnetism in $(Li_{0.8}Fe_{0.2})OHFeSe$. *Nature Mater.* **14**, 325–329 (2015).

13. Zhao, L. *et al.* Common electronic origin of superconductivity in (Li,Fe)OHFeSe bulk superconductor and single-layer $FeSe/SrTiO_3$ films. *Nat. Commun.* **7,** 10608 (2016).

14. Niu, X. H. *et al.* Surface electronic structure and isotropic superconducting gap in $(Li_{0.8}Fe_{0.2})OHFeSe$. *Phys. Rev. B* **92,** 060504 (2015).

15. Lei, B. *et al.* Anomalous Evolution of High-Temperature Superconduc-tivity from a Low-$T_c$ Phase Tuned by Carrier Concentration in FeSe Thin Flakes. *Phys. Rev. Lett.* **116**, 077002 (2016).

16. Zhang, P. *et al.* Observation of two distinct $d_{xz}/d_{yz}$ band splittings in FeSe. *Phys. Rev. B* **91,** 214503 (2015).

17. Richard, P. *et al.* Fe-based superconductors: an angle-resolved photo-emission spectroscopy perspective. *Rep. Prog. Phys.* **74,** 124512 (2011).

18. Huang, D. *et al.* Revealing the Empty-State Electronic Structure of Single-Unit-Cell $FeSe/SrTiO_3$. *Phys. Rev. Lett.* **115**, 017002 (2015).

19. Yeh, J. *et al.* Atomic Data and Nuclear Data Tables, **32**, 1 (1985).

20. Wu, X. X. *et al.* Topological characters in $Fe(Te_{1-x}Se_x)$ thin films. *Phys. Rev. B* **93,** 115129 (2016).

21. Liu, Z. H. *et al.* Three Dimensionality and Orbital Characters of the Fermi Surface in $(Tl, Rb)_yFe_{2-x}Se_2$. *Phys. Rev. Lett.* **109**, 037003 (2012).

22. Lee, J. J. *et al.* Interfacial mode coupling as the origin of the enhancement of $T_c$ in FeSe films on $SrTiO_3$. *Nature* **515**, 245–248 (2014).

23. Zhang, P. *et al.* Observation of high-Tc superconductivity in rectangular FeSe/STO(110) monolayer. Preprint at http://arxiv.org/abs/1512.01949 (2015).

24. Norman, M. R. *et al.* Phenomenology of the low-energy spectral function in high-$T_c$ superconductors. *Phys. Rev. B* **57**, R11093 (1998).

25. Zhang, W. H. *et al.* Effects of Surface Electron Doping and Substrate on the Superconductivity of Epitaxial FeSe Films. *Nano Lett.* 16, 1969−1973 (2016).

26. Tang, C. J. *et al.* Superconductivity dichotomy in K-coated single and double unit cell FeSe films on SrTiO₃. *Phys. Rev. B* **92**, 180507 (2015).

27. Richard, P. *et al.* ARPES measurements of the superconducting gap of Fe-based superconductors and their implications to the pairing mechanism. *J. Phys.: Condens. Matter* **27**, 293203 (2015).

28. Zhang, Y. *et al.* Superconducting gap anisotropy in monolayer FeSe thin film. Preprint at <http://arxiv.org/abs/1512.06322> (2015).

29. Nakayama, K. *et al.* Reconstruction of Band Structure Induced by Electronic Nematicity in an FeSe Superconductor. *Phys. Rev. Lett.* **113**, 237001 (2014).

30. Wu, X. X. *et al.* Nematic orders and nematicity-driven topological phase transition in FeSe. Preprint at <http://arxiv.org/abs/1603.02055> (2016).

## Acknowledgements

We thank Fu-Chun Zhang for useful discussions. This work is supported by grants from the Ministry of Science and Technology of China (2015CB921000, 2015CB921301) and the National Natural Science Foundation of China (11574371, 11274362, 1190020, 11334012, 11274381).

## Author contributions

Z.-Q.H., X.-L.P. and Y.-J.S. synthesized the samples. X.S., Z.-Q.H., M.-W.Q. and T.Q. performed the ARPES measurements. X.S. analysed the data. X.S., P.R., J.P.H., Y.-J.S. and H.D. wrote the manuscript. H.D. and Y.-J.S. supervised the project. All authors discussed the paper.

## Additional information

The authors declare no competing financial interests. Correspondence and requests for materials should be addressed to Y.-J.S. (yjsun@iphy.ac.cn) and H.D. (dingh@iphy.ac.cn).

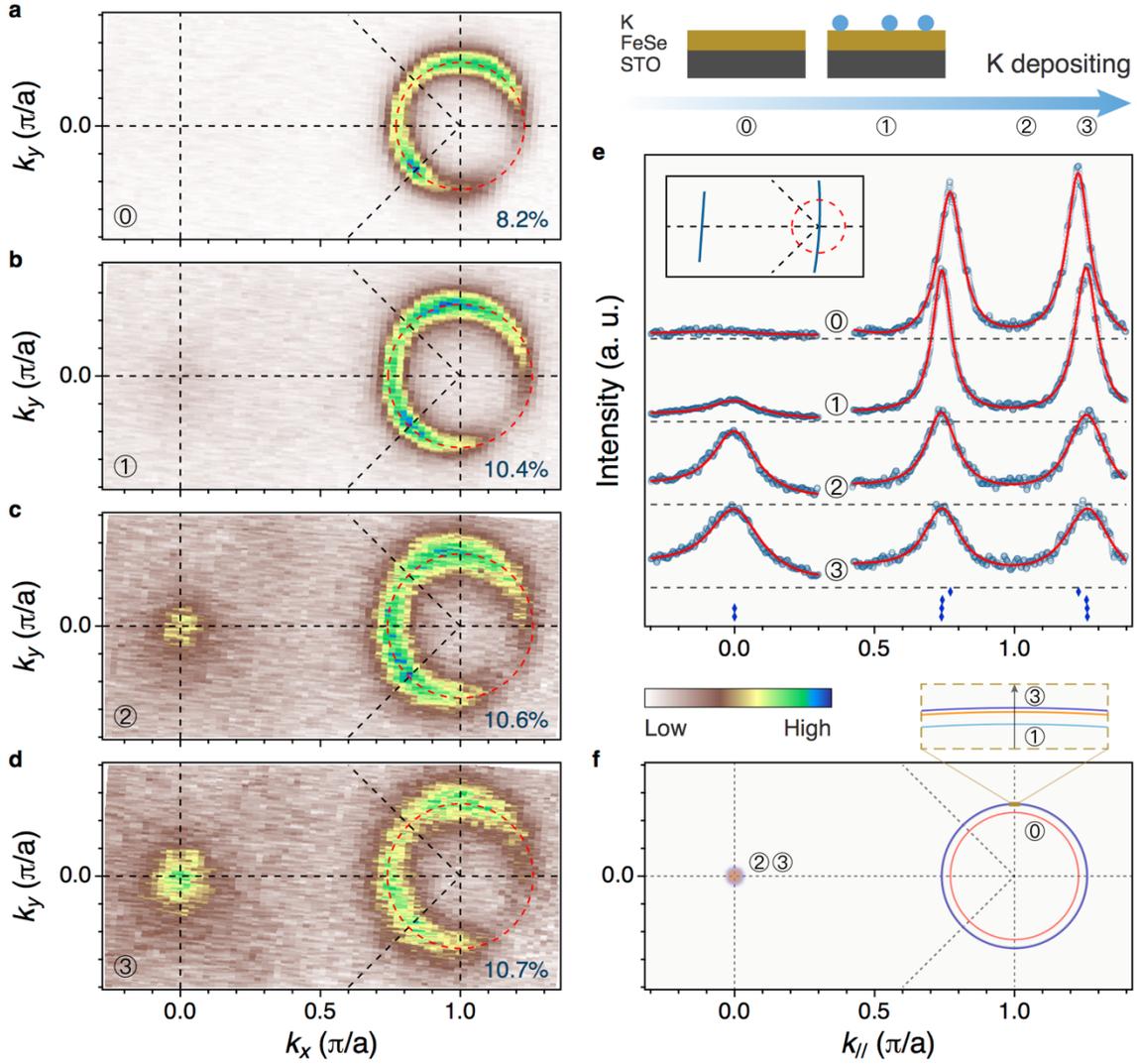

**Figure 1 | FS evolution of potassium coated 1UC FeSe/STO. a,** FS intensity map of a pristine sample recorded at 20 K and integrated within a 20 meV energy window with respect to $E_\text{F}$. The doubly-degenerate electron pocket occupies an area of $\sim$ 8.2% of the whole BZ, thus giving an electron carrier concentration of 0.164 electrons per unit cell according to the Luttinger theorem. **b-d,** Same as **a** but for the film after potassium was coated continuously. The percentages at the right bottom of each panel indicate the areas of the electron FS around M in the 1-Fe BZ. The size of the electron pocket at M is saturating slowly upon successive rounds of potassium deposition ($\sim$ 10.4%, $\sim$ 10.6% and $\sim$ 10.7% for the first, second and third rounds of deposition, respectively), but instead the spectrum gets broadened, due to the induced disorder at the surface. **e,** Evolution of the MDCs along the high-symmetry cuts

indicated in the inset upon potassium coating. The red curves correspond to fits of the data using multiple Lorentz functions. **f,** Comparison of the FSs shown in **a-d**.

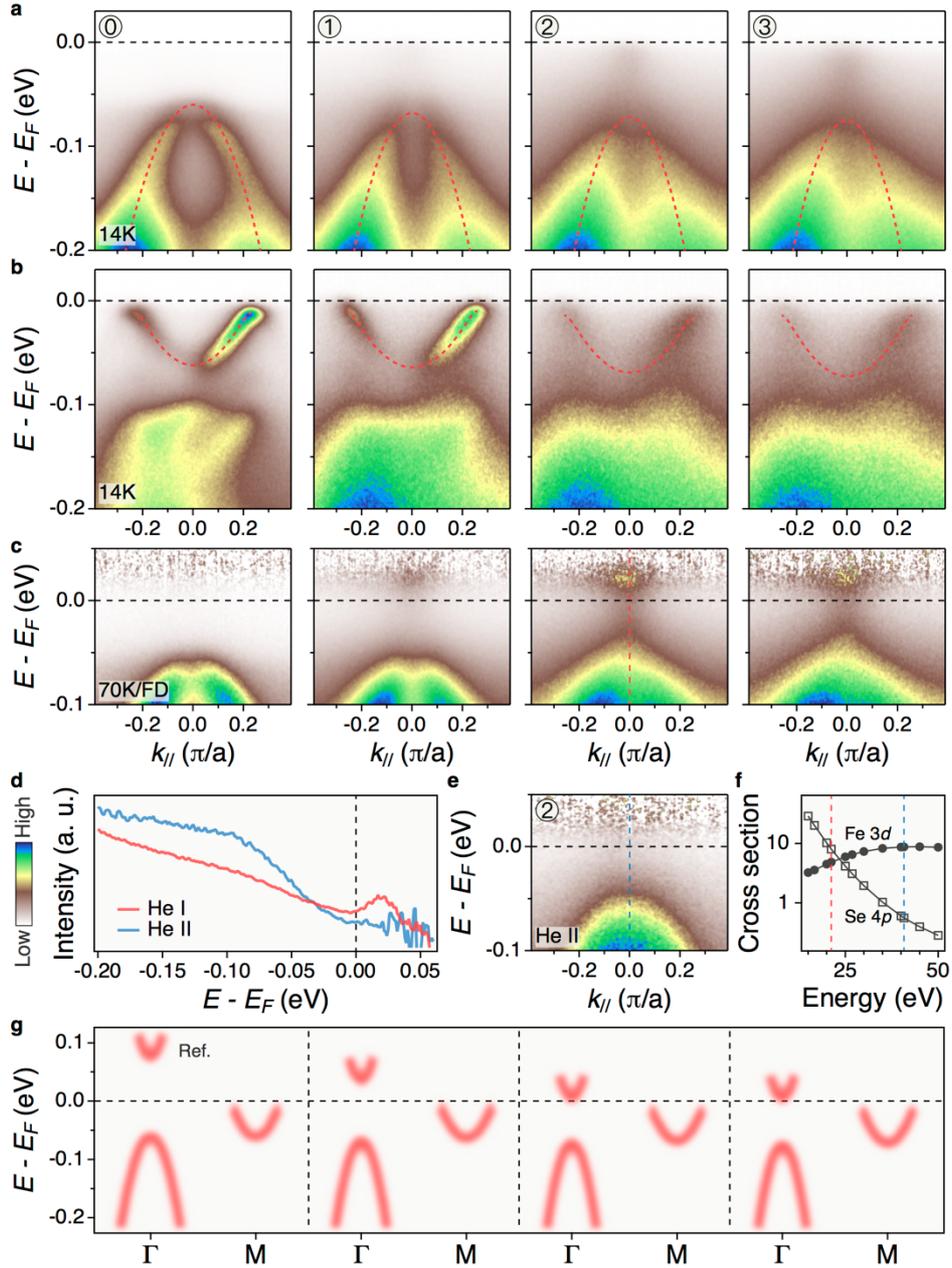

**Figure 2 | Electronic band structure. a-b,** Potassium coating evolution of the ARPES intensity plots at 14 K near Γ and M along the direction shown in the inset of Fig. 1 **f**. The dashed red curves are parabolic fits to the band dispersions. **c,** Intensity plots along the same cut as in **a**, but recorded at 70 K. The plots are divided by the Fermi-Dirac distribution function convoluted by the resolution function to visualize the states above $E_F$. **e,** Intensity plot near Γ recorded with He II rather than He Iα photons for the potassium coated sample labelled as 2. **d,** Comparison of the EDCs at Γ recorded with He Iα and He II beams. **f,** Calculated atomic photonionization cross sections for Fe 3*d* and Se 4*p*. **g,** Comparison of the

band dispersions along the Γ-M high-symmetry line. The energy positions of the electron-like band around Γ are taken from the reference or estimated from the data in **c**.

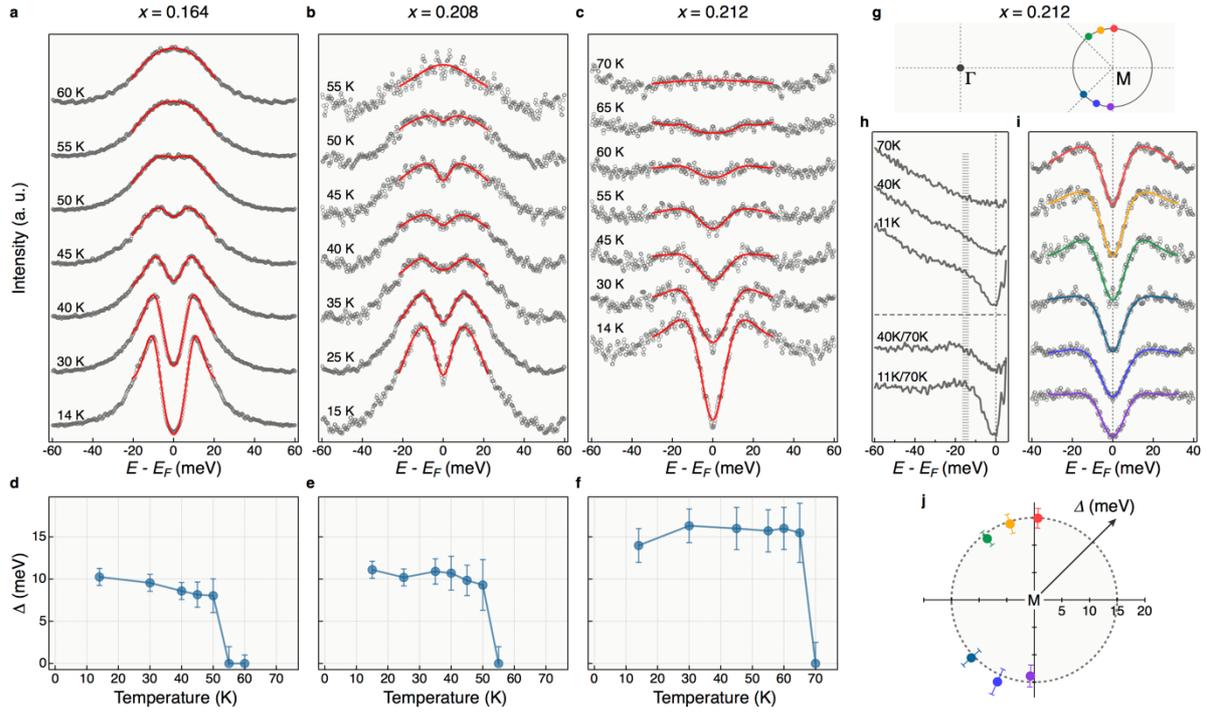

**Figure 3 | Superconducting gap. a-c,** Temperature evolution of the symmetrized EDCs at the $k_F$ point of the electron FS around M for pristine and potassium-coated 1UC FeSe/STO. The electron doping is indicated above the panels. The red curves correspond to fit of the data. **d-f,** Superconducting gap sizes as a function of temperature obtained from the fits shown in **a-c**, respectively. **g,** Schematic FS of K-coated 1UC FeSe/STO with doping $x \sim 0.212$. **h,** Temperature evolution of the EDCs at Γ divided by the FD function. **i,** Symmetrized EDCs at 14 K measured at various $k_F$ points as indicated by coloured dots in **g**. **j,** Polar representation of the momentum dependence of the superconducting gap size for the electron FS around M. A nearly-isotropic gap is highlighted by the dashed grey circle at 15 meV.

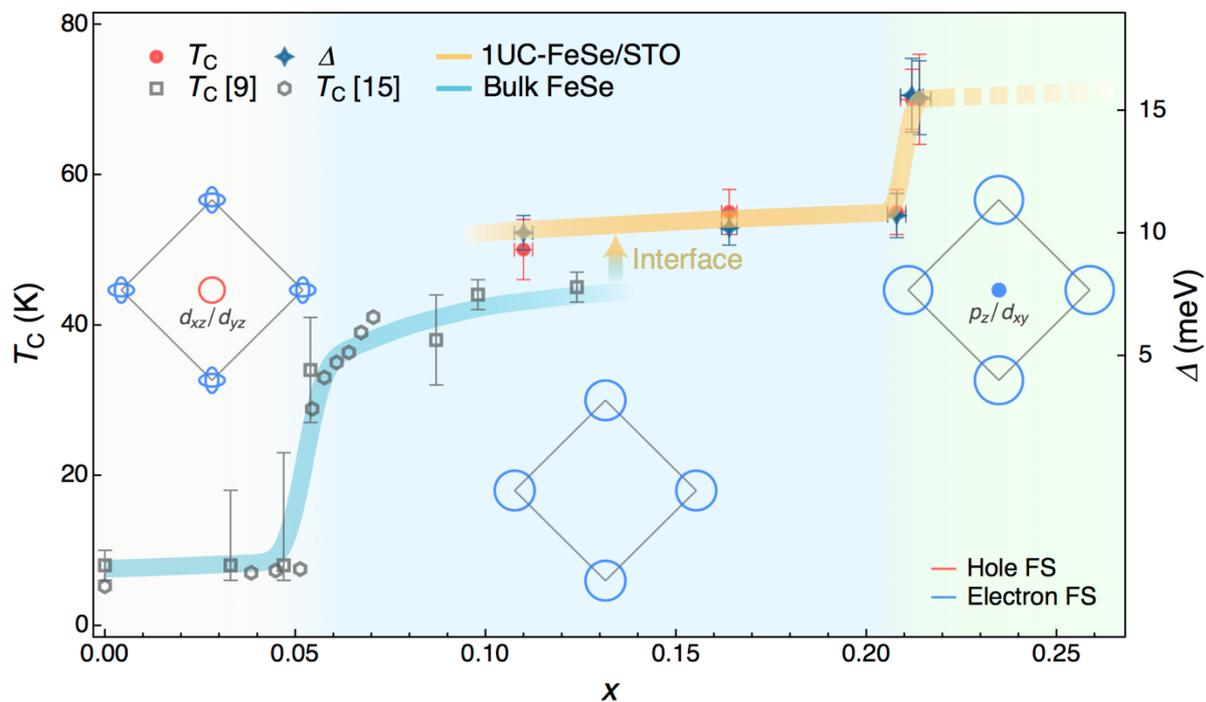

**Figure 4 | Schematic phase diagram of FeSe.** The phase diagram illustrates the evolution of superconductivity and the FS topology. The data at x = 0.11 is from our previous study [23]. The $T_c$ data of bulk FeSe, as traced with a cyan curve, is adapted from Refs. [9, 15].

**SUPPLEMENTARY**

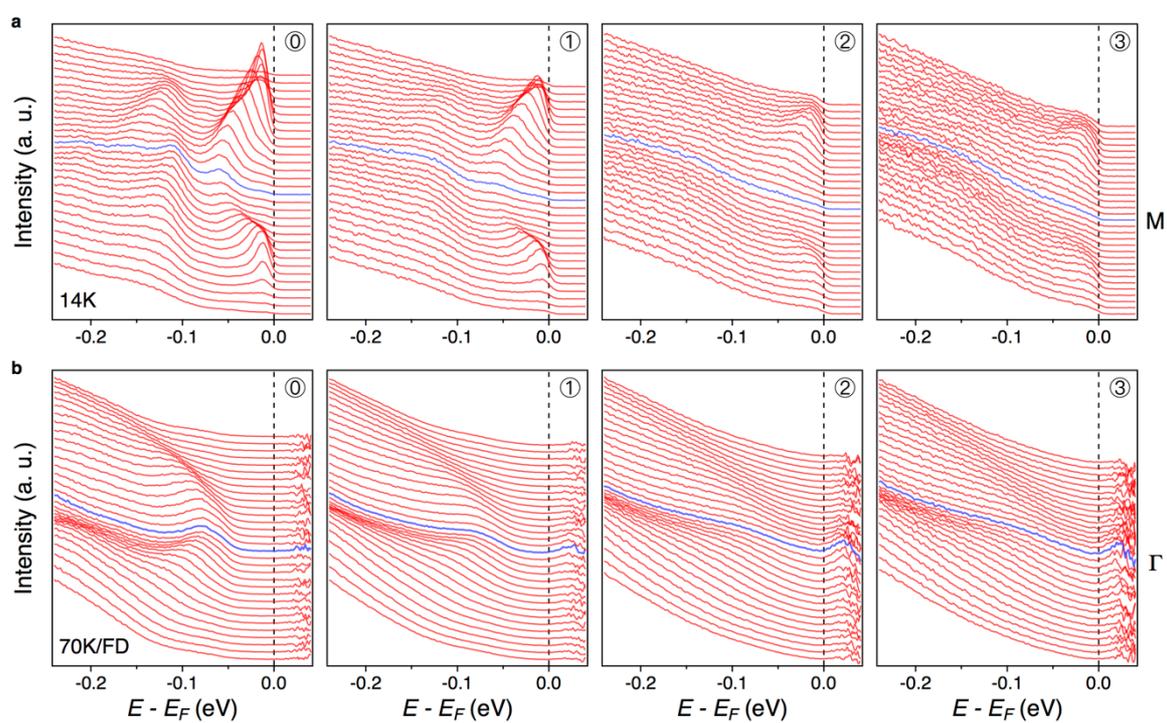

**Figure S1 | Evolution of the energy distribution curves (EDCs) upon K deposition. a,** EDC plots corresponding to the data in Fig. 2b. **b,** Same as **a** but corresponding to Fig. 2c.

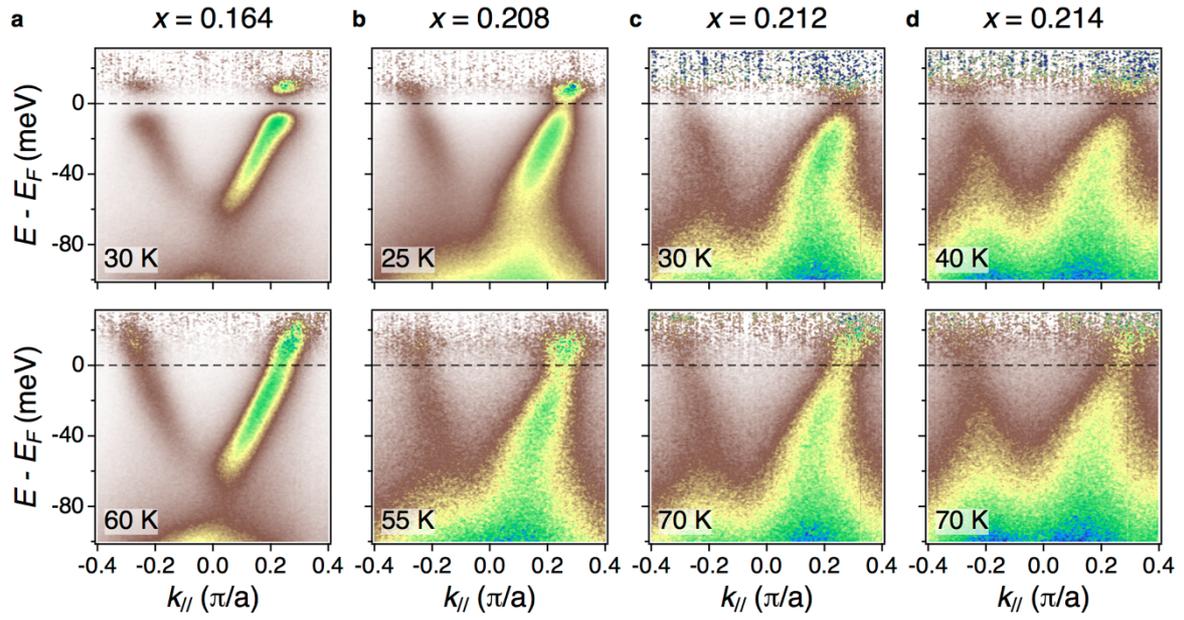

**Figure S2 | Comparison of the spectra at different temperatures for each sample. a,** ARPES intensity plots near the M point recorded at the low and high temperature for the pristine 1UC FeSe/STO. The spectra are divided by the Fermi-Dirac function in order to access partly the unoccupied states. **b-d,** Same as **a** but for the potassium-coated film with the electron-doping level indicated above the panels, respectively.

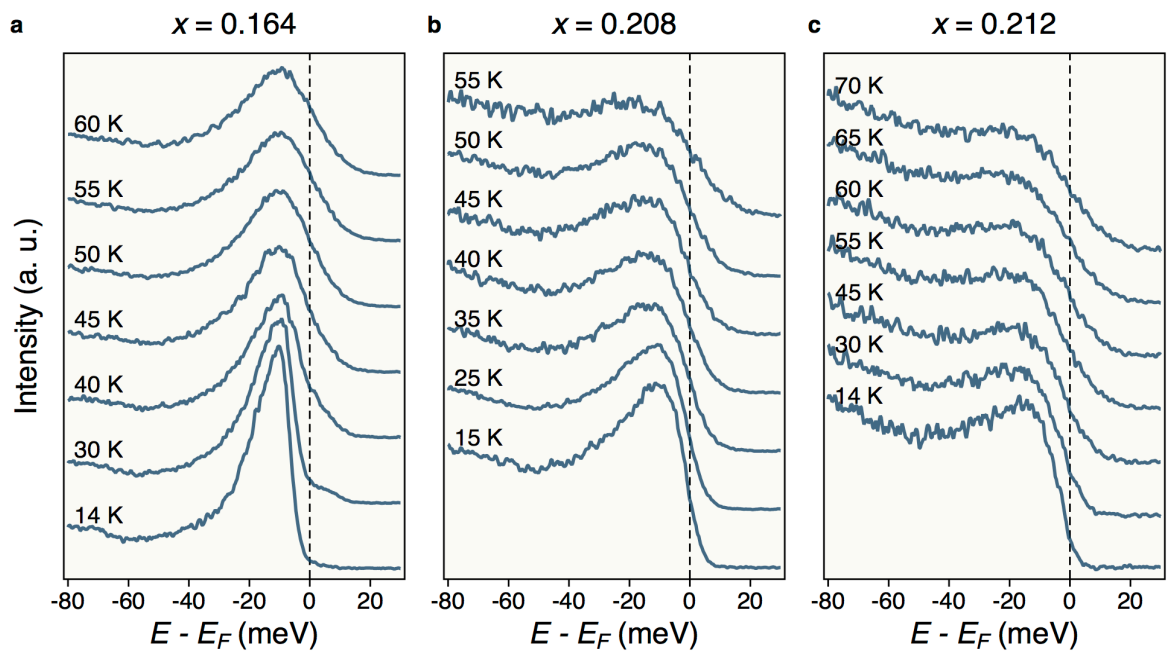

**Figure S3 | Temperature evolution of the EDCs at the $k_F$ point of the electron FS around M. a-c,** For pristine and potassium-coated 1UC FeSe/STO.

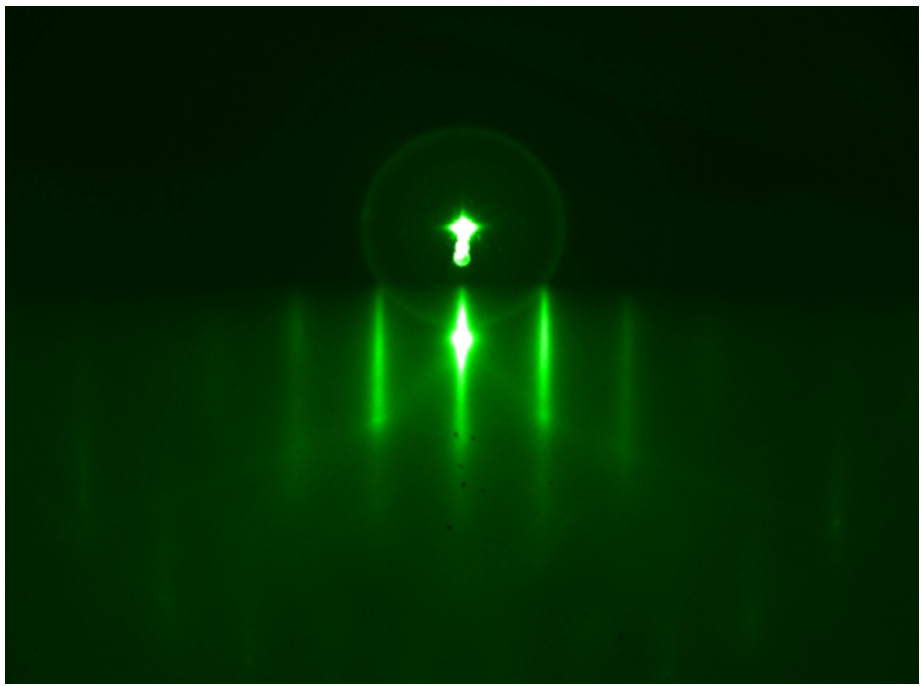

**Figure S4 | Refection high-energy electron diffraction (RHEED) image of the pristine 1UC FeSe/STO after annealing.**